# Improved Selective Harmonic Elimination for Reducing Torque Harmonics of Induction Motors in Wide DC Bus Voltage Variations

Hossein Valiyan Holagh, Tooraj Abbasian Najafabadi
School of Electrical and Computer Engineering
College of Engineering, University of Tehran
Tehran, Iran
h.valiyan.h@ut.ac.ir, najafabadi@ut.ac.ir

Mohsen Mahoor
Dept. of Electrical and Computer Engineering
University of Denver
Denver, CO, USA
Mohsen.Mahoor@du.edu

*Abstract*— Conventionally, Selective Harmonic Elimination (SHE) method in 2-level inverters, finds best switching angles to reach first voltage harmonic to reference level and eliminate other harmonics, simultaneously. Considering Induction Motor (IM) as the inverter load, and wide DC bus voltage variations, the inverter must operate in both over-modulation and linear modulation region. Main objective of the modified SHE is to reduce harmonic torques through finding the best switching angles. In this paper, optimization is based on optimizing phasor equations in which harmonic torques are calculated. The procedure of this method is that, first, the ratio of the same torque harmonics is estimated, secondly, by using that estimation, the ratio of voltage harmonics that generates homogeneous torques is calculated. For the estimation and the calculation of the ratios motor parameter, mechanical speed of the rotor, the applied frequency, and the concept of slip are used. The advantage of this approach is highlighted when mechanical load and DC bus voltage variations are taken into consideration. Simulation results are presented under a wide range of working conditions in an induction motor to demonstrate the effectiveness of the proposed method.

*Index Terms*--Induction motor drive, torque harmonics reduction, modified SHE PWM, various dc bus voltage, low switching frequency.

## I. Introduction

AC motor drives, operated with PWM inverters, produce pulsating torque due to non-sinusoidal nature of the inverter output voltage [1], [2]. Modulation methods have been discussed to reduce the pulsating torque in induction motor drives such as space-vector modulation [3], [4], Advanced PWM techniques [5], [6] and vector-control [6]. The fundamental of the voltages applied by inverters generates the average output torque, and the harmonic voltages generate increased losses and torque pulsations. The contribution of the harmonics to the average torque is not significant [7], [8].

There are different methods to calculate or estimate the harmonic currents, torques, and losses. These methods are presented in [7]–[11]. The losses are separated into various components and it is shown that the largest loss is usually happening in the rotor bars [7]. Harmonic losses are shown to be nearly independent of motor load and the fundamental magnetizing current is found to increase over that. In [11], the machine d-q models have been used to calculate the torque whereas the single-phase equivalent circuit is also used for studying the steady-state behavior of the machine [7]–[10]. Reference [8] shows that torque fluctuations are due mainly to the interaction of fundamental flux in the air gap at rotor current harmonics. In [10], the major torque ripple in the induction motor that is the 6th harmonic component has been estimated based on the fundamental harmonic and other harmonics of the rotor current.

In megawatt-rated inverter-fed induction motor drive systems, the switching frequency must be kept at a low value of only a few hundred hertz. Not only low switching frequency operation reduces the device switching losses, but also brings down the device dv/dt requirement. This issue leads to improve the inverter efficiency, and lesser Electro Magnetic Interference (EMI) generation from the inverter [18]. The reduction of switching frequency generally results in increasing current distortion, and consequently undesirable machine torque harmonics and higher machine losses. Compromises between excessive switching losses and undesirable harmonics can be achieved by employing optimal switching strategies for the inverter control. Off-line optimization methods shown in [12] are limited to drive systems with low dynamic requirements, and on-line optimized pulse control shown in [13] is suitable for high dynamic performance. The work in [14] shows the practical realization of a high dynamic performance and on-line optimized pulse control scheme based on field-orientation using a rectangular error boundary. Another PWM method by superposing a rectangular wave on the specific trapezoidal wave is presented to reduce torque ripple in induction motor [15]. Direct Self-Control (DSC) was mainly applied in high power applications, which required fast torque dynamic and low switching frequency. The study in [16] compares the stationary operational behavior of inverter-fed induction motor traction drives with high power and/or high dc-link voltage. Reference [17] proposed a new DTC method to reduce the torque ripple in five-phase IM. In [18], two optimal PWM schemes are proposed to minimize the low-order harmonic torque. The first approach which is independent of machine parameters and mechanical load is based on Frequency-Domain (FD), that is called classic SHE in this paper. The second approach considering the motor parameters and load torque is based on Synchronous Reference Frame (SRF). In this paper, a new control method is proposed to reduce the torque ripple in three-phase induction motor in low switching frequency. This approach introduced an on-line optimization method based on the parameters and the data of IM. Therefore, this approach constantly estimated the situation of the generated torque harmonics and determined the best switching angles to control. The effectiveness of the proposed approach is verified through MATLAB/SIMULINK simulation with an induction motor drive with a power equal to 3 kW.

The rest of this paper is organized as follows. In Section II, the summary of the torque harmonic calculations are provided and also their separations are accomplished. The proposed approach and the estimation of the phase difference and the amplitude ratios for the same torque harmonics are shown in section III. The calculation way of the switching angles and the evaluation of the maximum possible modulation index are presented in section IV. Section V provides the simulation results. The conclusion is presented in section VI.

## II. REVIEW AND ANALYSIS OF THE TORQUE HARMONICS

The effective voltage harmonics by switching in the three-phase induction motors are the fundamental voltage harmonic and the voltage harmonics of order ($6k\pm1$). Each of these harmonics produces a rotating magnetic field in the stator and the rotor of the induction machines. The fundamental voltage harmonic generates magnetic field with rotational speed $\omega_s$, the voltage harmonics of the order ($6k-1$) generate magnetic field with rotational speed $(-6k+1)\omega_s$, and the voltage harmonics of the order ($6k+1$) generate magnetic field with rotational speed $(6k+1)\omega_s$. The interaction between generated magnetic fields in the rotor and generated magnetic fields in the stator produce fixed and fluctuating electromagnetic torques. These fluctuating electromagnetic torques only included torque harmonics of the order $6k$. In other words, the voltage harmonics of the order ($6k\pm1$) manufacture the torque harmonics of the order $6k$ [7], [8], [19].

In this paper, single-phase equivalent circuit is used to calculate the ratio of amplitude and phase difference in the same torque harmonics. The effect of harmonics in the applied voltage into induction motors on the single-phase equivalent circuit parameters have been studied by in [7]. Single phase equivalent circuit is shown in Fig.1.

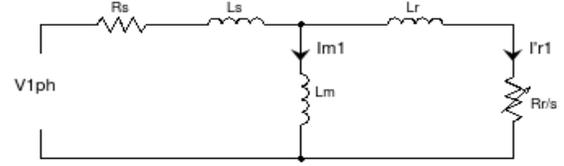

Figure 1. Single-phase equivalent circuits for induction motor.

A method for calculating the torque harmonics based on the single-phase equivalent circuit is shown in [7]–[10]. According to the method, torque fluctuations are engendered based on the interaction between fundamental air-gap flux harmonic and $(6k\pm1)^{th}$ current harmonics of the rotor in addition to the interaction between the fundamental current harmonic of the rotor and $(6k\pm1)^{th}$ air-gap flux harmonics. The torque harmonics generated based on input voltage harmonics can be distinguished as following:

$$\tau_{6k}^{6k-1} \propto \Psi_{m1}I'_{r(6k-1)}\sin(6k\omega_s t) + \Psi_{m(6k-1)}I'_{r1}\cos(6k\omega_s t) = A_{6k-1}\cos(6k\omega_s t + \theta_{6k-1})$$

$$\tau_{6k}^{6k+1} \propto -\Psi_{m1}I'_{r(6k+1)}\sin(6k\omega_s t) + \Psi_{m(6k+1)}I'_{r1}\cos(6k\omega_s t) = A_{6k+1}\cos(6k\omega_s t + \theta_{6k+1})$$
(1)

Where $\Psi_{m(6k-1)}$ and $\Psi_{m(6k+1)}$ are the $(6k-1)^{th}$ and $(6k+1)^{th}$ air-gap flux harmonics, respectively; $I'_{r(6k-1)}$ and $I'_{r(6k+1)}$ are the $(6k-1)^{th}$, and $(6k+1)^{th}$ current harmonics of the rotor; $\omega_s$ is the applied frequency to the stator; $\tau^{6k-1}$ and $\tau^{6k+1}$ are the torque harmonics generated from the $(6k-1)^{th}$ and $(6k+1)^{th}$ voltage harmonics respectively; $A_{6k\pm1}$ and $\theta_{6k\pm1}$ are the amplitude and the phase of the torque harmonics.

According to (1), torque harmonics can be formulated as following:

$$\tau_{6k} = A_{6k-1}\cos(6k\omega_s t + \theta_{6k-1}) + A_{6k+1}\cos(6k\omega_s t + \theta_{6k+1}) = A_{6k}\cos(6k\omega_s t + \theta_{6k})$$
(2)

where $\tau_{6k}$ is $6k^{th}$ torque harmonics, and $A_{6k}$ and $\theta_{6k}$ are their corresponding amplitude and phase, respectively.

Another method for calculating torque harmonics is presented in [4] which is based on the single-phase equivalent circuit. This method unlike the former has been presented approximately and only uses the rotor current harmonics for the estimation as following:

$$\tau_{6k} \cong \frac{3R_r}{2\omega_s} * \frac{1-s_1}{s_1} I'_{r1}(I'_{r(6k+1)} - I'_{r(6k-1)})\sin(6\omega_s t)$$
(3)

where $k = 1, 2, \ldots$, $S_1$ denotes the fundamental slip of the induction motor, and $R_r$ is the rotor resistance. The torque harmonics generated by fundamental and $(6k\pm1)^{th}$ harmonics of the current can be separated as following:

$$\tau_{6k-1} \propto -\frac{1-s_1}{\omega_s s_1} I'_{r1}I'_{r(6k-1)}\sin(6\omega_s t) \propto A_{6k-1}\cos(6k\omega_s t + \theta_{6k-1})$$
(4a)

$$\tau_{6k+1} \propto \frac{1-s_1}{\omega_s s_1} I'_{r1} I'_{r(6k+1)} \sin(6\omega_s t) \propto A_{6k+1} \cos(6k\omega_s t + \theta_{6k+1}) \quad (4b)$$

According to the above equations, the amplitude and the phase of $6k^{th}$ harmonics of the torque are proportional with the amplitude and phase of the fundamental and $(6k+1)^{th}$ harmonics of the rotor current. The single-phase equivalent circuit is used to calculate the amplitude and phase difference of the rotor currents. The current amplitude is proportional with applied voltage, frequency, and mechanical load, but the current phase is proportional only with applied voltage and frequency. According to fig.1 current phases are calculated as following:

$$\angle I'_{r1} - \angle V_{1ph} = \varphi_1 = \arctan\left(\frac{L_s + L_r}{(\frac{L_m}{L_m + L_S})^2 R_s + \frac{R_r}{s_1}}\right) \quad (5a)$$

$$\angle I'_{r(6k\pm1)} - \angle V_{(6k\pm1)ph} = \varphi_{6k\pm1} = \arctan\left(\frac{(6k\pm1)*(L_s + L_r)}{(\frac{L_m}{L_m + L_S})^2 R_s + \frac{R_r}{s_{6k\pm1}}}\right) \quad (5b)$$

Where $L_s$, $L_r$, and $L_m$ are stator, rotor and mutual inductances, respectively; $R_s$ is stator resistance, $S_{6k\pm1}$ are harmonic slips of the induction motor, and $\varphi$ is the phase difference between the rotor current and the corresponding voltage in every frequency. To simplify the calculations in (5a) and (5b), the voltage angles can be assumed equal to zero.

### III. THE PROPOSED METHOD FOR MINIMIZING TORQUE HARMONICS

In section II, in order to reduce the torque fluctuations, we separated the amplitude and the phase of the same generated torque harmonics. The phasor relationships of the equations are used to eliminate and minimize these fluctuations. To calculate the sum of two or more sinusoidal signals with the same frequency we can use their corresponding phasors instead of trigonometric expansions as following:

$$B_1 \cos(\omega t + \Delta\theta) + B_2 \cos(\omega t + \pi) = B_3 \cos(\omega t + \theta_3) \quad (6a)$$

where

$$(B_3)^2 = (B_1 \cos(\Delta\theta) - B_2)^2 + (B_1 \sin(\Delta\theta))^2 = f(B_1, B_2, \Delta\theta) \quad (6b)$$

$$\theta_3 = \arctan(\frac{B_1 \sin(\Delta\theta)}{B_1 \cos(\Delta\theta) - B_2}) \quad (6c)$$

To minimize the phasor phrases, (6b) should be minimized. For this purpose, (7) is used to reduce the amplitude of two signals of the same frequency.

$$\cos(\Delta\theta) = \frac{\min(B_1, B_2)}{\max(B_1, B_2)} \quad (7)$$

According to the expressed relationships in terms of the phasors, two general methods are proposed to decrease the amplitude of the torque harmonics. Two methods express that the possible minimum of the torque harmonic can be reached if the ratios of the same torque harmonics correspond the cosine of their frequency differences. In the first method, we have assumed that the amplitude of the $(6k-1)^{th}$ harmonics are always further than the amplitude of the $(6k+1)^{th}$ harmonics, but in the second method the amplitude of the $(6k+1)^{th}$ harmonics are further than the amplitude of the $(6k-1)^{th}$ harmonics. To achieve the proposed methods, the phase difference and the amplitude of the same torque harmonics must be estimated or calculated.

#### A. Estimation of the phase difference of the same torque harmonics

The expressed approximate equation in (3) is used to estimate the phase difference. According to (4a) and (4b), the phases of the same torque harmonics are equivalent with the corresponding current harmonic phases and the fundamental current harmonic phase. In this case, the phase differences of the same torque harmonics will be equal to the difference of two determined phases.

$$\Delta\theta_{6k} = \theta_{6k-1} - \theta_{6k+1} \cong 2\varphi_1 + \varphi_{6k-1} - \varphi_{6k+1} \quad (8)$$

According to (5b), the $\varphi_{6k\pm1}$ amounts are almost fixed for the certain $\omega_s$ when IM works from no-load to full load, however $\varphi_1$ is various and the $\varphi_1$ variation causes the variation of the phase difference of the same torque harmonics. The $\varphi_1$ variation is shown in Fig.2. It should be mentioned that the estimated relations are indefeasible with changing $\omega_s$.

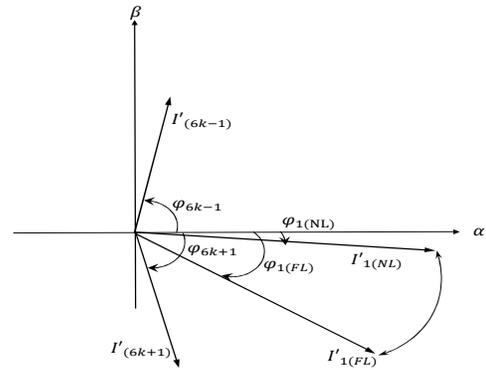

Figure 2. The variation of the fundamental current harmonic phase with changing the mechanical load from no-load to full load at the fixed synchronous speed.

#### B. Estimation of the ratios of the same torque harmonics

To estimate the ratios of the same torque harmonics, we use the concept of slip. Harmonic slips in the induction motors are defined as following:

$$s_{6k\pm1} = \frac{(6k\pm1)\omega_s \pm p\omega_m}{(6k\pm1)\omega_s} \quad (9)$$

where $\omega_m$ and $P$ are mechanical angular speed and number of pole pairs respectively. With changing the induction motor operation from no-load to full load, harmonic slips change correspondingly. This slip expresses the impact of voltage harmonics on their corresponding torque harmonics. According to (9) having two velocities, the mechanical speed and the synchronous speed, we can obtain the impact of the voltage harmonics on their corresponding torque harmonics.

The ratios of the same torque harmonics have the direct relation with the ratios of the voltage harmonics and the inverse relation have with the ratios of their corresponding harmonic slips.

$$\frac{A_{(6k-1)}}{A_{(6k+1)}} = \frac{V_{(6k-1)ph}}{V_{(6k+1)ph}} * \frac{s_{6k+1}}{s_{6k-1}} \quad (10)$$

With considering (7), (8), and (10), the ratio of two voltage harmonics which generate the same torque harmonics is defined in accordance with (11) to both proposed methods. This relation also for all the input frequency from no-load to maximum torque are confirmed, and a condition of using the expressed relations is that the IM works in its stable area.

$$\frac{V_{(6k-1)ph}}{V_{(6k+1)ph}} = \frac{s_{6k-1}}{s_{6k+1}} \cos(\Delta\theta_{6k}) \quad (11a)$$

$$\frac{V_{(6k+1)ph}}{V_{(6k-1)ph}} = \frac{s_{6k+1}}{s_{6k-1}} \cos(\Delta\theta_{6k}) \quad (11b)$$

## IV. CALCULATION OF THE BEST SWITCHING ANGLES

The proposed approach is suitable to control the induction motor torque when the number of the switching angles are very few, say k. However, in this paper we focus on two switching angles in each cycle. In the selective harmonic elimination method which employs two switching angles, the amplitude of fundamental harmonic is controlled and the 5$^{th}$ harmonic is eliminated. However, in this method the amplitude of the fundamental, 5$^{th}$, and 7$^{th}$ harmonics is controlled altogether. As a result of using two switching angles, the main torque is controlled and the 6$^{th}$ torque harmonic is minimized. On the other hand, when three switching angles are employed, the main torque is also controlled and the 6$^{th}$ and 12$^{th}$ torque harmonics are minimized. Fig. 3 illustrates the output voltage waveforms of voltage source inverter in which two switching angles are applied.

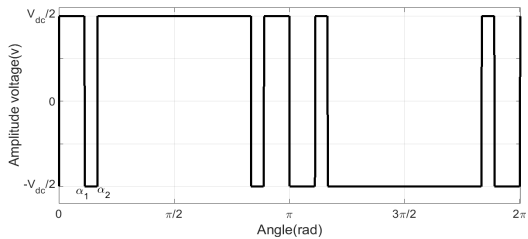

Figure 3. The output voltage waveforms of a voltage source inverter with two switching angles.

### A. Calculating the switching angles to minimize the 6$^{th}$ torque harmonic

In both proposed methods using two switching angles, the fundamental, 5$^{th}$, and 7$^{th}$ harmonics are controlled with the purpose of minimizing the 6$^{th}$ torque harmonic. According to Fig. 3, for the voltage waveform generated by voltage source inverter, the harmonic amplitudes on Fourier coefficients are calculated as following:

$$V_1 / (\frac{2}{\pi} V_{dc}) = MI = 1 - 2\cos(\alpha_1) + 2\cos(\alpha_2)$$

$$V_5 = (\frac{2V_{dc}}{5\pi})(1 - 2\cos(5\alpha_1) + 2\cos(5\alpha_2)) \quad (12)$$

$$V_7 = (\frac{2V_{dc}}{7\pi})(1 - 2\cos(7\alpha_1) + 2\cos(7\alpha_2))$$

where $\alpha_1$ and $\alpha_2$ are the switching angles, $MI$ is the modulation index, and $V_{dc}$ is the dc bus voltage. Based on (12), the switching angles are obtained from solving the following equations:

$$MI = 1 - 2\cos(\alpha_1) + 2\cos(\alpha_2)$$

$$\frac{V_5}{V_7} = (\frac{7}{5}) \frac{(1 - 2\cos(5\alpha_1) + 2\cos(5\alpha_2))}{(1 - 2\cos(7\alpha_1) + 2\cos(7\alpha_2))} = \frac{s_5}{s_7}\cos(\Delta\theta_6) \quad (I) \quad (13)$$

$$\frac{V_7}{V_5} = (\frac{5}{7}) \frac{(1 - 2\cos(7\alpha_1) + 2\cos(7\alpha_2))}{(1 - 2\cos(5\alpha_1) + 2\cos(5\alpha_2))} = \frac{s_7}{s_5}\cos(\Delta\theta_6) \quad (II)$$

$$0 \leq \alpha_1 < \alpha_2 \leq \pi/2$$

The slip is variable from zero to maximum torque slip. According to these change ranges, switching angles is calculated.

### B. Evaluation of the maximum modulation index in the change ranges

At this stage, due to the importance of the modulation index, we are surveying the maximum of modulation index that can be reached against the changes of the ratios of the voltage harmonics. Behavior of the modulation index changing is different in the two suggested methods. The curve of the maximum of modulation index extracted from (13) is shown in Fig.4. According to this curve, in the method (11a). The maximum possible amount of MI is 0.955, in the operation of maximum torque. On the other hand, in the method (11b), the maximum possible amount of modulation index is approximately equal to 1, in the operation of intermediate load. As a result, to reach the maximum modulation index using two switching angles, the ratio of the 7$^{th}$ voltage harmonic per the 5$^{th}$ voltage harmonic must be approached to about 0.7.

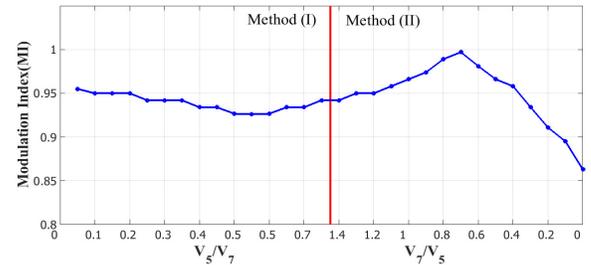

Figure 4. The curve of the maximum modulation index against the changes of the voltage harmonic ratio in the methods (11a) and (11b).

## V. SIMULATION RESULTS

For simulation, the three-phase squirrel-cage induction motor is used with the nominal information 3 kW, 380 V, and 1415 rpm. The parameters of this induction motor are listed in Table I. The induction motor is connected to a voltage source inverter that feed from a dc bus voltage that is equal to 560 V. In this paper, open loop control is considered as the type of control strategy. In simulations, our methods compare with two other methods. These methods are selective harmonic elimination (SHE-PWM), classic method, and two proposed methods. The suggested methods for the torque harmonic reduction are online. In other words, the presented methods need the ratios of voltage harmonics of the same torque based on the function of IM to calculate switching angles, furthermore the need to the modulation index. However, two first methods, SHE and classic SHE, using only the modulation index are able to calculate the switching angles.

TABLE I. INDOCTION MOTOR PARAMETERS.

| Parameter | Value |
|---|---|
| Rotor Resistance($R_s$) | 1.85 Ω |
| Rotor Resistance($R_r$) | 1.84 Ω |
| Stator Self Inductance($L_s$) | 170mH |
| Rotor Self Inductance($L_r$) | 170mH |
| Magnetizing Inductance ($L_m$) | 160mH |
| Moment of Inertia | 0.007 kg.m$^2$ |
| Number of Pole Pairs | 2 |

To evaluate the mentioned methods, three different working conditions for the induction motor torque are considered. In the first condition, the motor works under no-load, and the applied frequency to the IM is constant and equal to 50Hz. In the second condition, the IM is applied the linear load changing according to the mechanical speed of the rotor, and the frequency with which the motor works is equal to 50Hz. In the third condition, in addition to changing the applied linear mechanical load, the applied frequency is propounded equal to 45Hz until we can evaluate the proposed methods in the other applied frequency and the variable mechanical load.

*A. Evaluation of the Sixth Torque Harmonic Reduction in the First Condition*

The simulation results for four mentioned methods are shown in Fig. 5 when the IM works under no-load. Fig. 5 portrays the curves of the sixth torque harmonic amplitudes versus the modulation index in the four mentioned methods. In all methods, for each modulation index value, the motor slip is about zero.

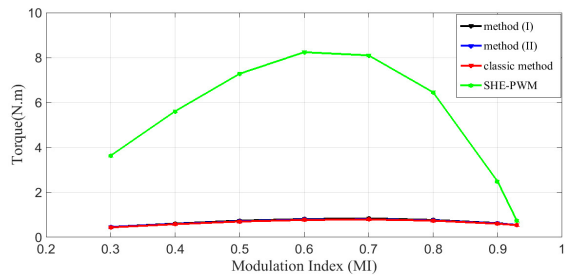

Figure 5. The amplitude of the sixth torque harmonics in the four mentioned methods under no-load in terms of the MI.

According to the curve of the selective harmonic elimination method in the first condition, we can conclude that the sixth harmonic amplitude firstly increased and secondly decreased with increasing the modulation index. In the other methods this result is indefeasible, but their variations are less. According to the numerical amounts, the torque harmonic amplitudes for the all modulation range in the classic, (11a), and (11b) methods are lesser than SHE-PWM method. And the numerical value of the torques in three methods, classic, (11a), and (11b), are almost equal because the ratio the fifth voltage harmonic per the seventh voltage harmonic is constant in the whole range of modulation. In the first Condition for reducing the sixth torque harmonic, the classic method is superior to the other methods because that needs less memory than two methods (11a) and (11b) whereas these two methods due to being online need more memory.

*B. Evaluation of the Sixth Torque Harmonic Reduction in the second Condition*

In the second condition, the linear load in terms of the mechanical speed and the frequency equal to 50Hz applied to the IM. Fig. 6 shows the simulation results for the second condition.

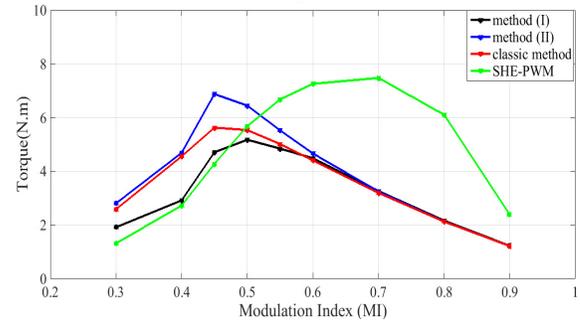

Figure 6. The amplitude of the sixth torque harmonics in the four mentioned methods under linear load in terms of the MI at the applied frequency equal to 50Hz.

According to the obtained amounts, the sixth torque harmonics is firstly increased and secondly decreased with increasing the modulation index in four mentioned methods. In the second condition, because the IM is under linear load, the motor slip is changing from specified amount to about zero with increasing the modulation index. The motor slip changes conclude the phasor situation variation of sixth torque harmonic produced with the 5$^{th}$ and 7$^{th}$ voltage harmonics. According to the numerical values of the torque harmonic amplitudes, the appropriate method selection for reducing the torque fluctuations is different with changing the modulation index or the motor slip. In the first condition, because the motor slip is about zero, the classic approach is more appropriate than the other techniques. In this case, when the motor slip is close to zero, the classic method and two methods (11a) and (11b) have better performance like the first condition. Moreover, when the motor slip amount reaches the maximum value which maximum electromagnetic torque is produced, SHE-PWM method and method (11a) are appropriate approaches for reducing torque fluctuations. Finally, the proposed method (11a) is suitable when the IM works in the middle of the slip, between the maximum slip and about zero. It should be noted that motor operation area is more in the middle area of the motor slip, which depends on the mechanical load. Thus method (11a) is suitable way in the whole motor operation area such as under no-load, middle-load, and full-load. Furthermore, the operation of method (11b) in the middle area of the slip is better than SHE-PWM method, and another advantage of this method compared to other methods is that it can be reached the maximum modulation index.

## C. Evaluation of the Sixth Torque Harmonic Reduction in the third Condition

In the third condition, in line with changing the applied linear load, we consider the applied frequency equal to 45Hz until we can evaluate the proposed methods in the variable frequency and load. Due to the applied linear load, the modulation index can change from a specified value to a peak value. The simulation results for four mentioned methods are shown in fig. 7. According to the obtained numerical results, all the results expressed in the second condition can also be concluded. In this condition with changing the applied frequency, the proposed methods still have their previous performance.

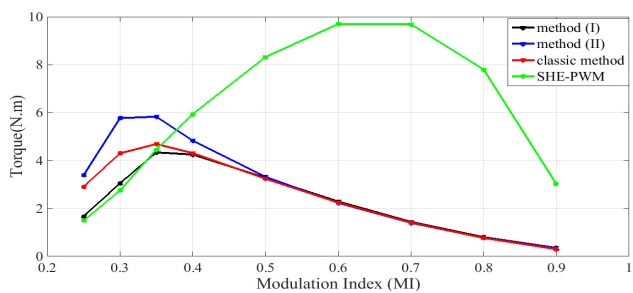

Figure 7. The amplitude of the sixth torque harmonics in the four mentioned methods under linear load in terms of the MI at the applied frequency equal to 45Hz.

## VI. CONCLUSION

The classic method for the torque fluctuates reduction is suitable when either the IM is under no-load or the motor slip is negligible. In fact, this method is not proper when the motor slip is at its middle area or at its maximum value. SHE-PWM approach operates better only when the IM works at the maximum possible slip. Accordingly, each of the available approaches only operates at the specific area of working range. Whereas, the proposed approach in this paper provided the best function in the entire working range of the IM as this method constantly detects the position of the generated torque harmonics, and then selects the best switching angles, in order to minimize the fluctuations. This approach was divided into the two methods. In the first method, It was assumed that the amplitude of the $(6k-1)^{th}$ torque harmonics were always further than the amplitude of the $(6k+1)^{th}$ torque harmonics, but this was conversely in the second method The considered assumption was compatible with the inverter operations, so that the first method presented the best function among other methods. In addition, the second method can only be reached the maximum modulation index. Finally, as an extension of the approach, we suggested implementation of the presented approach on SVM-PWM working in over- modulation region.

REFERENCES

[1] W. Liang, W. Fei, and P. C.-K. Luk, "Analytical investigation of sideband torque ripple in induction machine drive with SPWM and SVPWM techniques," *International Conference on Electrical Machines and Systems (ICEMS), 2014 17th*, 2014, pp. 162-168.

[2] J. Song-Manguelle, G. Ekemb, S. Schröder, T. Geyer, J.-M. Nyobe-Yome, and R. Wamkeue, "Analytical expression of pulsating torque harmonics due to PWM drives," *Energy Conversion Congress and Exposition (ECCE), 2013 IEEE*, 2013, pp. 2813-2820.

[3] B. Singh, S. Jain, and S. Dwivedi, "Torque ripple reduction technique with improved flux response for a direct torque control induction motor drive," *IET Power Electronics,* vol. 6, pp. 326-342, 2013.

[4] U. V. Patil, H. M. Suryawanshi, and M. M. Renge, "Closed-loop hybrid direct torque control for medium voltage induction motor drive for performance improvement," *IET Power Electronics,* vol. 7, pp. 31-40, 2014.

[5] K. Basu, J. S. Prasad, G. Narayanan, H. K. Krishnamurthy, and R. Ayyanar, "Reduction of torque ripple in induction motor drives using an advanced hybrid PWM technique," *IEEE Transactions on Industrial Electronics,* vol. 57, pp. 2085-2091, 2010.

[6] V. P. K. Hari and G. Narayanan, "Space-vector-based hybrid PWM technique to reduce peak-to-peak torque ripple in induction motor drives," *IEEE Transactions on Industry Applications,* vol. 52, pp. 1489-1499, 2016.

[7] E. A. Klingshirn and H. E. Jordan, "Poly-phase induction motor performance and losses on nonsinusoidal voltage sources," *IEEE Trans. on power apparatus and systems,* vol. PAS-87, pp. 624-631, Mar. 1968.

[8] S. D. Robertson and K. Hebbar, "Torque pulsations in induction motors with inverter drives," *IEEE Trans. on Industry and General Applications,* vol. IGA-7, pp. 318-323, Mar./Appr. 1971.

[9] G. Jainy, "The effect of voltage waveshape on the performance of a 3-phase induction motor," *IEEE Trans. on Power Apparatus and Systems,* vol. 83, pp. 561-566, 1964.

[10] B. K. Bose, Power Electronics and AC Drives. Englewood Cliffs, NJ: Prentice-Hall, 1986.

[11] P. C. Krause, "Method of multiple reference frames applied to the analysis of symmetrical induction machinery," *IEEE Trans. on Power Apparatus and Systems,* pp. 218-227, Jan. 1968.

[12] G. S. Buja and G. B. Indri, "Optimal pulsewidth modulation for feeding AC motors," *IEEE Trans. Ind. Appl. , Vol. A1-13,* pp. 38-44, Jan./Febr. 1977.

[13] J. Holtz and S. Stadtfeld, "A predictive controller for the stator current vector of ac machines fed from a switched voltage source," in *JIEE IPEC-Tokyo Conf*, 1983, pp. 1665-1675.

[14] A. Khambadkone and J. Holtz, "Low switching frequency and high dynamic pulsewidth modulation based on field-orientation for high-power inverter drive," *IEEE trans. power elec. ,* vol. 7, pp. 627-632, Oct. 1992.

[15] K. Taniguchi, M. Inoue, Y. Takeda, and S. Morimoto, "A PWM strategy for reducing torque-ripple in inverter-fed induction motor," *IEEE tran. Ind. Appl. ,* Vol. 30, pp. 71-77, Jan./Febr. 1994.

[16] A. Steimel, "Direct self-control and synchronous pulse techniques for high-power traction inverters in comparison," *IEEE Trans. on Ind. Elec. ,* vol. 51, pp. 810-20, Aug. 2004.

[17] Y. N. Tatte and M. V. Aware, "Torque ripple reduction in five-phase direct torque controlled induction motor," in *Power Electronics, Drives and Energy Systems (PEDES), 2014 IEEE International Conference on*, 2014, pp. 1-5

[18] A. Tripathi and G. Narayanan, "Evaluation and Minimization of Low-Order Harmonic Torque in Low-Switching-Frequency Inverter-Fed Induction Motor Drives," *IEEE Trans. Ind. Appl. ,* vol. 52, pp. 1477-1488, 2016.